\documentstyle[fancybox,epsfig,color,axodraw,12pt]{article}

\def\plb#1{Phys.~Lett.~{\bf B#1}}
\def\npb#1{Nucl.~Phys.~{\bf B#1}}
\def\prl#1{Phys.~Rev.~Lett.~{\bf #1}}
\def\prd#1{Phys.~Rev.~{\bf D#1}}

\def\mt{{\ifmmode M^{eff}_T\else $M^{eff}_T$\fi}}

\def\e{\epsilon}

\def\L{\lambda}

\def\e3{$\epsilon_3$}

\def\ch2{$\chi^2$}

\def\co#1{{\ifmmode{\cal O}_{#1}\else${\cal O}_{#1}$\fi}}

\def\dltmh{$\Delta m_H^2 \;\;$}
\def\dltmb{$\Delta m_b$}

\newdimen\unit
\def\point#1 #2 #3{\vbox to0pt{\kern-#2\unit
  \hbox{\kern#1\unit#3}\vss}
 \nointerlineskip}

\newcommand{\be}{\begin{equation}}
\newcommand{\ee}{\end{equation}}
\newcommand{\bea}{\begin{eqnarray}}
\newcommand{\eea}{\end{eqnarray}}

\begin{document}
\thispagestyle{empty} \noindent
\begin{flushright}
        October 2002
\end{flushright}

\vspace{1cm}
\begin{center}
Higgs and SUSY Particle Predictions from SO(10) Yukawa
Unification\footnote{Talk given at the 1st International
Conference on String Phenomenology, Oxford, UK, July 6 - 11,
2002.}
\end{center}
  \vspace{1cm}
    \begin{center}
S. Raby\\
      \vspace{0.3cm}
\begin{it}
Department of Physics, The Ohio State University, \\ 174 W. 18th
Ave., Columbus, Ohio  43210
\end{it}
  \end{center}
  \vspace{1cm}
\centerline{\bf Abstract}
\begin{quotation}
\noindent  In this talk we assume SO(10) boundary conditions at
the GUT scale, including unification for the third generation
Yukawa couplings $\lambda_t = \lambda_b = \lambda_\tau$.  We find
that this assumption is only consistent with the low energy data
in a narrow region of soft SUSY breaking parameter space.  We
discuss the consequences of this result for Higgs and SUSY
searches.
\end{quotation}
\vfill\eject

\section{ SUSY GUTs }

\subsection{Soft SUSY Breaking Parameters and ``Naturalness"}\label{subsec:search}

Supersymmetric particles have still not been discovered.  Many
supersymmetry [SUSY] enthusiasts are becoming discouraged.
``Naturalness" constraints suggest a spectrum of light SUSY
particles with mass of order a few hundred GeV,  IF we demand fine
tuning less than 1 in 10.  On the other hand, if we allow for fine
tuning of order 1 in 1000, then SUSY particles with mass of order
a TeV are fine.  How much fine tuning is too much?   Recall that
in the standard model, the problem we are trying to solve (why the
Higgs is so much lighter than, say, the GUT scale) requires fine
tuning to 1 part in $10^{28}$ for the Higgs mass squared. Perhaps
1 part in 1000 is not so bad.

Another guide for SUSY searches comes by assuming that the LSP
provides the observed dark matter in the universe.  Using such
arguments, several authors (see for example, Ellis and Nanopoulos
in these proceedings) have obtained ``natural" ranges for soft
SUSY breaking parameters.

In this talk we discuss a different guide for SUSY
searches.\footnote{This talk is based on two papers in
collaboration with T. Bla\v zek and R. Derm\' \i \v sek
\cite{bdr}.}  We show that SO(10) boundary conditions at the GUT
scale, for soft SUSY breaking parameters as well as for the Yukawa
couplings of the third generation, are consistent with the low
energy data, including $M_t, \ m_b(m_b), \ M_\tau$, ONLY in a
narrow region of SUSY breaking parameter space.   Moreover, this
region is also preferred by constraints from CP and flavor
violation, as well as by the non-observation of proton decay.
Finally we discuss the consequences for the Higgs and SUSY
spectrum.

\subsection{Virtues of SO(10) SUSY GUT}
 Supersymmetric grand unified theories have many virtues.
 Supersymmetry alone provides a framework for solving the gauge hierarchy problem
and a mechanism for naturally obtaining electroweak symmetry
breaking with a heavy top quark.   In addition, GUTs explain the
charge assignments of quarks and leptons, i.e. charge quantization
\cite{guts}.

Recall that in ${\bf SU_5}$ the quarks and leptons of one family
are described by $ \{ Q = \left(\begin{array}{c} u
\\ d \end{array}\right) \;\;\; \bar {e}
\;\;\; \bar{u} \} \;\; \subset \;\; \bf 10$  and $  \{ {\bar{d} \;\;\; L} = \left(\begin{array}{c} \nu \\
 e \end{array}\right) \} \;\; \subset \;\; \bf \bar{5}$.   And the
 two Higgs doublets are given by
$ H_u ,\;\; H_d \;\; \subset \;\; {\bf 5_H},\;\; {\bf \bar{5}_H}$.

 In ${\bf SO_{10}}$ we have the more compelling unification of all quarks and leptons of one
 family into one irreducible representation such that  $ {\bf 10} + {\bf \bar{5}} +
 \bar{\nu}_{sterile} \;\; \subset \;\; { \bf 16}$ and the two
 Higgs doublets are also unified with ${\bf 5_H},\;\; {\bf \bar{5}_H} \;\; \subset \;\;
{\bf 10_H} $.

Moreover at the moment the only experimental evidence for
supersymmetry is through the successful prediction of gauge
coupling unification\cite{gaugeunif,gaugetwo,gaugethree}. This
prediction is now tested at the level of two loop renormalization
group running from the GUT to the weak scales. Self-consistency
thus requires including one loop threshold corrections at both the
weak and GUT scales. It is important to note that there are
significant GUT threshold corrections from the Higgs and GUT
breaking sectors.   It is thus useful to define the GUT scale
$M_G$ as the scale where $\alpha_1(M_G) = \alpha_2(M_G) \equiv
\tilde \alpha_G$.  A good fit to the low energy data then requires
a threshold correction $\epsilon_3 \equiv \frac{(\alpha_3(M_G) -
\tilde \alpha_G)}{\tilde \alpha_G} \sim - 4\%$.

\subsection{$SO_{10}$ Yukawa unification}

Minimal $SO_{10}$ also predicts Yukawa unification for the third
family of quarks and leptons with $\lambda_b = \lambda_t =
\lambda_{\tau} = \lambda_{\nu_\tau} = \lambda$ at the GUT
scale\cite{yukunif}.

Ignoring threshold corrections, one can use the low energy value
for $m_b/m_{\tau}$ to fix the universal Yukawa coupling $\lambda$.
RG running from $M_G$ to $M_Z$ then gives $\lambda_\tau(M_Z)$.
Hence given $m_\tau = \lambda_\tau \frac{v}{\sqrt{2}} cos\beta$ we
obtain $\tan\beta \approx 50$. Finally, a prediction for the top
quark mass is given by $m_t = \lambda_t \frac{v}{\sqrt{2}}
sin\beta \sim 170 \pm 20 \; {\rm GeV}$ (see Anderson et
al.\cite{yukunif}).

Note, in this case there are insignificant GUT threshold
corrections from gauge and Higgs loops.  Nevertheless, the
previous discussion is essentially a {\it straw man}, since there
are huge threshold corrections at the weak scale\cite{threshcorr}.
The dominant contributions are from gluino and chargino loops plus
an overall logarithmic contribution due to finite wave function
renormalization given by $\delta m_b/ m_b = \Delta m_b^{\tilde g}
+ \Delta m_b^{\tilde \chi} + \Delta m_b^{\log} + \cdots$.   These
contributions are characteristically of the form \begin{equation}
\Delta m_b^{\tilde g} \approx \frac{2 \alpha_3}{3 \pi} \;
\frac{\mu m_{\tilde g}}{m_{\tilde b}^2} \; tan\beta  ,
\label{eq:gluino} \end{equation}
 \begin{equation} \Delta m_b^{\tilde \chi^+} \approx  \frac{\lambda_t^2}{16 \pi^2}
\; \frac{\mu A_t}{m_{\tilde t}^2} \; tan\beta  \;\;\; {\rm and}
\label{eq:chargino} \end{equation}
\begin{equation}  \Delta m_b^{\log} \approx \frac{\alpha_3}{4
\pi} \log(\frac{\tilde m^2}{M_Z^2}) \sim 6 \% \label{eq:log}
\end{equation} with $\Delta m_b^{\tilde g} \sim - \Delta
m_b^{\tilde \chi} > 0$ {\em for {\bf $\mu > 0$}} [with our
conventions].   These corrections can easily be of order $\sim 50$
\%.   However good fits require\ $\delta m_b/ m_b <  - 2\%$.

Note, the data favors $\mu > 0$.  First consider the process
{\large $b \rightarrow s \gamma$}.  The chargino loop contribution
typically dominates and has opposite sign to the standard model
and charged Higgs contributions for $\mu > 0$, thus reducing the
branching ratio.   This is desirable, since the standard model
contribution is a little too large.  Hence $\mu < 0$ is
problematic when trying to fit the data.  Secondly, the recent
measurement of the anomalous magnetic moment of the muon suggests
a contribution due to NEW physics given by $a_\mu^{NEW} = 26 (16)
\times 10^{-10}$\cite{gminus2}. However in SUSY the sign of
$a_\mu^{NEW}$ is correlated with sign of $\mu$ \cite{nath}. Once
again the data favors $\mu > 0$.

Before discussing our analysis of Yukawa unification, we need to
consider one important point.  $SO(10)$ Yukawa unification with
the minimal Higgs sector necessarily predicts large $\tan\beta
\sim 50$.   It is much easier to obtain EWSB with large
$\tan\beta$ when the Higgs up/down masses are split ($m_{H_u}^2 <
m_{H_d}^2$)~\cite{ewsb}. In our analysis we consider two
particular Higgs splitting schemes we refer to as Just So and D
term splitting. In the first case the third generation squark and
slepton soft masses are given by the universal mass parameter
$m_{16}$, and only Higgs masses are split:  $m_{(H_u, \; H_d)}^2 =
m_{10}^2 \;( 1 \mp \Delta m_H^2)$. In the second case we assume D
term splitting, i.e. that the D term for $U(1)_X$ is non-zero,
where $U(1)_X$ is obtained in the decomposition of $SO(10)
\rightarrow SU(5) \times U(1)_X$.  In this second case, we have $
m_{(H_u,\; H_d)}^2 = m_{10}^2 \mp 2 D_X $, $m_{(Q,\; \bar u,\;
\bar e)}^2 = m_{16}^2 + D_X$,  $m_{(\bar d,\; L)}^2 =  m_{16}^2 -
3 D_X $. The Just So case does not at first sight appear to be
very well motivated. However we now argue that it is quite
natural~\cite{bdr}.  In $SO(10)$, neutrinos necessarily have a
Yukawa term coupling active neutrinos to the ``sterile" neutrinos
present in the {\bf 16}. In fact for $\nu_\tau$ we have
$\L_{\nu_\tau} \; \bar \nu_\tau \; L \; H_u$ with $\L_{\nu_\tau} =
\L_t = \L_b = \L_\tau \equiv \; {\bf \L}$. In order to obtain a
tau neutrino with mass $m_{\nu_\tau} \sim 0.05$ eV (consistent
with atmospheric neutrino oscillations), the ``sterile" $\bar
\nu_\tau$ must obtain a Majorana mass $M_{\bar \nu_\tau} \geq
10^{13}$ GeV.   Moreover, since neutrinos couple to $H_u$ (and not
to $H_d$) with a fairly large Yukawa coupling (of order 0.7), they
naturally distinguish the two Higgs multiplets. With $\L = 0.7$
and $M_{\bar \nu_\tau} = 10^{13}$ GeV, we obtain a significant GUT
scale threshold correction with $\Delta m_H^2 \approx 10$\%,
remarkably close to the value needed to fit the data.   At the
same time, we obtain a small threshold correction to Yukawa
unification $\approx 2.5$\%.

\subsection{$\chi^2$ Analysis}

Our analysis is a top-down approach with 11 input parameters,
defined at $M_G$, varied to minimize a $\chi^2$ function composed
of 9 low energy observables. The 11 input parameters are: $M_G, \;
\alpha_G(M_G),$ $ \epsilon_3$; the Yukawa coupling $\L$, and the 7
soft SUSY breaking parameters $\mu,\; M_{1/2},\; A_0, \;
\tan\beta$, $m_{16}^2, \; m_{10}^2$,  \dltmh
 ($D_X $) for Just So (D term) case.
We use two (one)loop renormalization group [RG] running for
dimensionless (dimensionful) parameters from $M_G$ to $M_Z$ and
complete one loop threshold corrections at
$M_Z$~\cite{pierceetal}.  We require electroweak symmetry breaking
using an improved Higgs potential, including $m_t^4$ and $m_b^4$
corrections in an effective 2 Higgs doublet model below
$M_{stop}$~\cite{carenaetal}.  Note, in the figures we have chosen
to keep three input parameters $\mu,\; M_{1/2},\; m_{16}$ fixed,
minimizing $\chi^2$ with respect to the remaining 8 parameters
only.
   The \ch2 function includes the 9 observables; 6 precision electroweak data $\alpha_{EM},\;
G_\mu, \;  \alpha_s(M_Z) = 0.118 \ (0.002),\; M_Z, \; M_W, \;
\rho_{NEW}$ and the 3 fermion masses $M_{top} = 174.3 \ (5.1),\;
m_b(m_b) = 4.20 \ (0.20), \; M_\tau$.

\begin{figure}[th]
\centerline{\epsfxsize=12cm\epsfbox{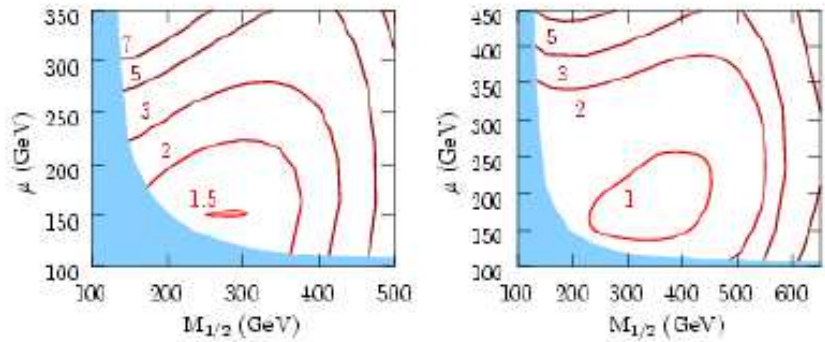}}

\caption{$\chi^2$ contours for $m_{16} = 1500$ GeV (Left) and
$m_{16} = 2000$ GeV (Right). The shaded region is excluded by the
chargino mass limit $m_{\tilde \chi^+} > 103$ GeV. \label{chi2}}
\end{figure}

Fig. 1 (Left) shows the constant $\chi^2$ contours for $m_{16} =
1500$ GeV in the case of Just So squark and slepton masses. We
find acceptable fits ($\chi^2 < 3$) for $A_0 \sim - 1.9 \;
m_{16}$, $m_{10} \sim 1.4 \; m_{16}$ and $ m_{16} \geq 1.2$ TeV.
The best fits are for $m_{16} \geq 2000$ GeV with $\chi^2 < 1$.
Fig. 1 (Right) shows the constant $\chi^2$ contours for $m_{16} =
2000$ GeV.

\begin{figure}[th]
\centerline{\epsfxsize=12cm\epsfbox{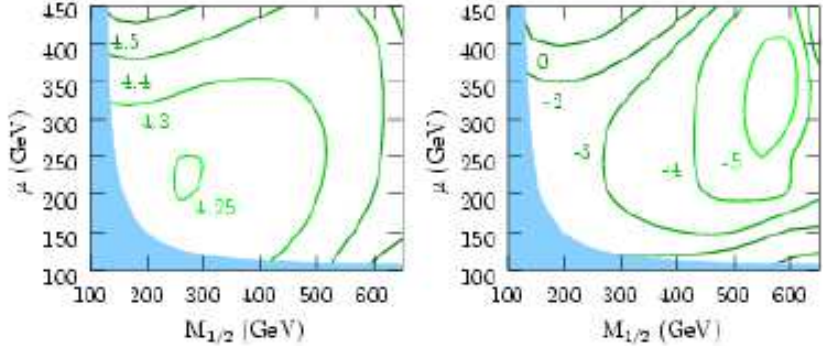}}

\caption{Contours of constant $m_b(m_b)$[GeV] (Left) and \dltmb in
\% (Right) for $m_{16} = 2000$ GeV. \label{m_b_and_corr}}
\end{figure}

Fig. 2 gives the constant $m_b(m_b)$ and $\delta m_b/ m_b$
contours for $m_{16} = 2000$ GeV. We see that the best fits, near
the central value, are found with $\delta m_b/ m_b \leq - 2$\%.
The chargino contribution (Eqn. \ref{eq:chargino}) is typically
opposite in sign to the gluino (Eqn. \ref{eq:gluino}), since $A_t$
runs to an infrared fixed point $\propto - M_{1/2}$(see for
example, Carena et al.\cite{threshcorr}).  Hence in order to
cancel the positive contribution of both the log (Eqn.
\ref{eq:log}) and gluino contributions, a large negative chargino
contribution is needed. This can be accomplished for $- A_t >
m_{\tilde g}$ and $m_{\tilde t_1} << m_{\tilde b_1}$. The first
condition can be satisfied for $A_0$ large and negative, which
helps pull $A_t$ away from its infrared fixed point.   The second
condition is also aided by large $A_t$. However in order to obtain
a large enough splitting between $m_{\tilde t_1}$ and $m_{\tilde
b_1}$, large values of $m_{16}$ are needed.    Note, that for Just
So scalar masses, the lightest stop is typically lighter than the
sbottom. We typically find $m_{\tilde b_1} \sim 3 \; m_{\tilde
t_1}$. On the other hand, D term splitting with $D_X
> 0$ gives $m_{\tilde b_1} \leq m_{\tilde t_1}$.   As a result in
the case of Just So boundary conditions excellent fits are
obtained for top, bottom and tau masses; while for D term
splitting the best fits give $m_b(m_b) \geq 4.59$ GeV.

The bottom line is that Yukawa unification is only possible in a
narrow region of SUSY parameter space with \begin{equation} A_0
\sim - 1.9 \;  m_{16}, \;\; m_{10} \sim 1.4 \; m_{16},\;\;\;
\end{equation}
\begin{equation} (\mu,\ M_{1/2}) \sim 100 - 500 \; {\rm GeV \;\; and} \;\;  m_{16}
\geq 1.2 \; {\rm TeV} . \end{equation} It would be nice to have
some a priori reason for the fundamental SUSY breaking mechanism
to give these soft SUSY breaking parameters. However, without such
an a priori explanation, it is all the more interesting and
encouraging to recognize two additional reasons for wanting to be
in this narrow region of parameter space.

\subsection{Inverted Mass Hierarchy \& Proton Decay Bounds}

One mechanism for suppressing large flavor violating processes in
SUSY theories is to demand heavy first and second generation
squarks and sleptons (with mass $\gg$ TeV) and the third
generation scalars lighter than a TeV.  Since the third generation
scalars couple most strongly to the Higgs, this limit can still
leave a ``naturally" light Higgs.  It was shown that this inverted
scalar mass hierarchy can be obtained via renormalization group
running from $M_G$ to $M_Z$ with suitably chosen soft SUSY
breaking boundary conditions at $M_G$ \cite{scrunching}.  All that
is needed is $SO(10)$ boundary conditions for the Higgs mass (i.e.
$m_{10}$), squark and slepton masses (i.e. $m_{16}$) and a
universal scalar coupling $A_0$. In addition, they must be in the
ratio
$$A_0^2 = 2 \ m_{10}^2 = 4 \ m_{16}^2, \;\;\; {\rm with} \;\;\;
m_{16} \gg \; {\rm TeV}.$$

Secondly, in order to suppress the rate for proton decay due to
dimension 5 operators one must also demand \cite{pdecay} $$
(\mu,\; M_{1/2}) << m_{16}, \;\;\; {\rm with} \;\;\; m_{16} > \
{\rm few \; TeV} .$$

\section{Consequences for Higgs and SUSY Searches}

In Fig. 3 we show the constant light Higgs mass contours
 for $m_{16} = 1500$ and $2000$ GeV
(solid lines) with the constant $\chi^2$ contours overlayed
(dotted lines).  Yukawa unification for $\chi^2 \leq 1$  clearly
prefers a light Higgs with mass in a narrow range,  112 -  118
GeV.

\begin{figure}[th]
\centerline{\epsfxsize=12cm\epsfbox{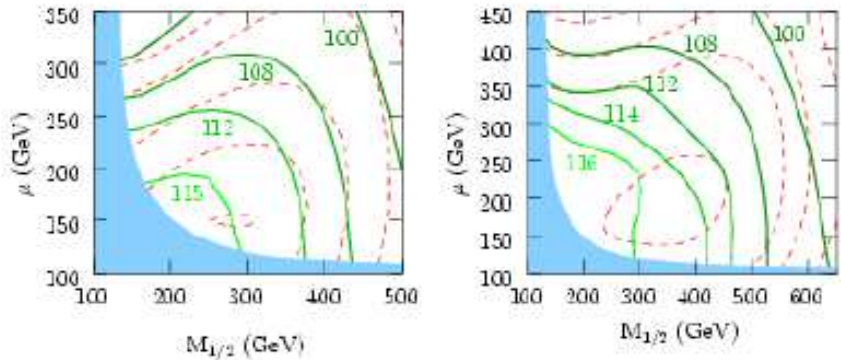}}

\caption{Contours of constant $m_h$ [GeV] (solid lines) with
$\chi^2$ contours from Fig. 1 (dotted lines) for $m_{16} = 1500$
GeV (Left) and $m_{16} = 2000$ GeV (Right). \label{h0_and_chi2}}
\end{figure}

In this region the CP odd $A^0$, the heavy CP even Higgs $H^0$ and
the charged Higgs bosons $H^\pm$ are also quite light.  In
addition we find the mass of $\tilde t_1 \sim (150 - 250)$ GeV,
$\tilde b_1 \sim (450 - 650)$ GeV, $\tilde \tau_1 \sim (200 -
500)$ GeV, $\tilde g \sim (600 - 1200)$ GeV, $\tilde \chi^+ \sim
(100 - 250)$ GeV, and $\tilde \chi^0 \sim (80 - 170)$ GeV.  All
first and second generation squarks and sleptons have mass of
order $m_{16}$.  The light stop  and chargino may be visible at
the Tevatron.  With this spectrum we expect $\tilde t_1
\rightarrow \tilde \chi^+ \;b $ with $\tilde \chi^+ \rightarrow
\tilde \chi^0_1 \; \bar l \; \nu$ to be dominant. Lastly $\tilde
\chi^0_1$ is the LSP and possibly a good dark matter
candidate~\cite{leszek}.

Our analysis thus far has only included third generation Yukawa
couplings; hence no flavor mixing.   If we now include the second
family and 2-3 family mixing, consistent with $V_{c b}$, we obtain
new and significant constraints on $m_{\tilde t_1}$ and $m_{A^0}$.
The stop mass is constrained by $B(b \rightarrow s \gamma)$ to
satisfy $m_{\tilde t}^{MIN} > 450$ GeV (unfortunately increasing
the bottom quark mass). In addition, as shown by Babu and Kolda
\cite{bsmumu} the one loop SUSY corrections to CKM mixing angles
(see Bla\v{z}ek et al.\cite{threshcorr}) result in flavor
violating neutral Higgs couplings. As a consequence the CDF bound
on the process $B_s \rightarrow \mu^+ \mu^-$ places a lower bound
on $m_{A^0} \geq 200$ GeV \cite{bsmumu}.  \ch2, on the other hand,
increases as $m_{A^0}$ increases.   However the increase in \ch2
is less than 60\% for $m_{A^0} < 400$ GeV.  Note, the $H^\pm, \
H^0$ masses increase linearly with $m_{A^0}$.

In conclusion, we have demanded $SO(10)$ Yukawa unification for
the third generation and, instead of predicting the top, bottom
and tau masses, we have turned the tables around and used it to
predict Higgs and SUSY particle masses.  We have shown that Yukawa
unification only works in a narrow region of soft SUSY breaking
parameters. This same region is also preferred (1) for suppressing
large SUSY CP and flavor violation with an inverted scalar mass
hierarchy and (2) suppressing proton decay due to dimension 5
operators.  We find a SUSY particle spectrum with light gauginos,
third generation squarks and sleptons lighter than a TeV, but
first and second generation scalars heavier than a TeV.  We find
$m_h^0 \sim  114 \pm 5 \pm 3$ GeV where the first uncertainty
comes from the range of SUSY parameters with \ch2 $\leq 1.5$ and
the second is an estimate of the theoretical uncertainties in our
Higgs mass.  The light Higgs mass is naturally in this range as a
consequence of having large $\tan\beta$ and a light stop.  Since
we necessarily have $m_{16} > 1200$ GeV, we obtain a small SUSY
contribution to the anomalous magnetic moment of the muon with
$a_\mu^{SUSY} < 16 \times 10^{-10}$.  Finally,  our best results
are obtained with a light CP odd Higgs.  However the CDF bound on
the process $B_s \rightarrow \mu^+ \mu^-$ places a lower bound on
$m_{A^0} \geq 200$ GeV \cite{bsmumu}.   We would thus not be
surprised to see evidence for $B_s \rightarrow \mu^+ \mu^-$ in Run
II at the Tevatron.

\end{document}